\documentclass[fleqn,usenatbib]{mnras}
\usepackage{cancel}

\usepackage{newtxtext,newtxmath}

\usepackage{tabularx}
\usepackage{amssymb}
\usepackage{amsfonts} 
\usepackage{amsmath}
\usepackage{comment}
\usepackage{float}

\usepackage[T1]{fontenc}
\usepackage{wasysym}

\DeclareRobustCommand{\VAN}[3]{#2}
\let\VANthebibliography\thebibliography
\def\thebibliography{\DeclareRobustCommand{\VAN}[3]{##3}\VANthebibliography}
\newcommand{\argmin}{\arg\!\min}

\usepackage{graphicx}	
\usepackage{amsmath}	
\usepackage{amssymb}	
\usepackage{float}
\usepackage{multirow}
\usepackage{multicol}

\title[Shape Reconstruction of Transiting Celestial Bodies]
{{\it COD:} An Algorithm for Shape Reconstruction of Transiting Celestial Bodies through Topological Optimization}

\author[Nachmani, Mazeh and Sochen]{
G. Nachmani$^{1,2}$,  
T. Mazeh$^{2}$,
N.Sochen$^{1}$
 \\
$^{1}$Department of Applied Mathematics, Tel Aviv University, Tel Aviv 69978, Israel\\
$^{2}$ School of Physics and Astronomy, Faculty of Exact Sciences, Tel Aviv University, Tel Aviv 69978, Israel\\
}

\date{Accepted 2022 January 11. Received 2021 December 27; in original form 2021 October 27}

\pubyear{2022}

\begin{document}
\label{firstpage}
\pagerange{\pageref{firstpage}--\pageref{lastpage}}
\maketitle

\begin{abstract}
We introduce a novel algorithm, \textit{COD} --- Compact Opacity Distribution, for shape reconstruction of a celestial body that has been observed to occult a star, using the photometric time-series observations of the occultation. 
\textit{COD} finds a solution to the light-curve inversion problem for an optically thick occulter having an approximately convex shape, together with an estimate of its size, impact parameter and velocity, relative to the occulted star.
The algorithm is based on an optimization scheme that uses topological constraints and an objective function for the geometry of the occulter.
The constraints of the problem follow linear relations, which enable the use of linear programming 
optimization as the mathematical framework.
Multiple tests of the algorithm were performed, all of which
resulted in high 
correlations between the simulated and obtained shapes of the occulting objects, with errors within $5\%$ in their projected velocities and horizontal sizes, and within $0.1$ in their impact parameters. These tests include a video of a solar eclipse by Phobos, as seen by NASA's Curiosity rover, which was collapsed into its corresponding light curve and reconstructed afterwards.
We applied \textit{COD} to the mysterious case of VVV-WIT-08 --- a single deep occultation (\( \sim 96 \% \)) 
of a giant star 
lasting for over 200 days.
The analysis, which did not assume any specific shape of the occulter, suggested an object with a projected opacity distribution resembling an ellipse with an eccentricity of \( \sim 0.5 \), tilted at \( \sim 30 \) degrees relative to the direction of motion, 
with a semi-minor axis similar to the stellar radius.


\end{abstract}


\section{Introduction}
\label{section:Introduction}
\addcontentsline{toc}{section}{Introduction}

In recent years, extensive photometric surveys of millions of stars, obtained by ground and space observations, revealed multiple large dips in stellar light curves, which due to their shapes, are unlikely to have originated from spherical occulters, such as planets or secondary stars \citep[e.g.,][]{Boyajian16, Ansdell16, Rappaport19}.
Since the angular size of main-sequence stars is significantly smaller than the diffraction limit of current optical telescopes, it is still impossible to resolve neither the star nor the shape of the eclipsing body. Therefore, only a scalar value of the total amount of stellar light blocked by the transiting body at each instant in time is measured in that case.

The observed dips, which lasted between hours and hundreds of days and blocked up to $\sim 96\%$ of the stellar flux, were presumably caused by unknown transiting bodies, such as planetary ring systems \citep{Mamajek12, Bourne18}, warped inner-disc edge or cometary-like objects \citep{Scaringi16}, a family of large comets or planetesimals \citep{Boyajian16,Bodman16}, dust clouds or transient accretion events of dust \citep{Rappaport19}, clumps of circumstellar material \citep{Ansdell16}, unseen companions with circumsecondary disks \citep{Osborn17}, an asymmetric dusty disk \citep{Kloppenborg15} and possibly sublimating icy moons \citep{Bourne18}. 

Though there may be clues about the astrophysical nature of the occulters, through the temporal behaviour of their transit light curves or their spectral characteristics, for example, their specific shapes generally remain unknown.
The standard approach
for the reconstruction of unknown shapes of occulting bodies
uses a model-based solution, which assumes a specific shape of the transiting body with a small number of free parameters, which are determined by curve fitting. 
For example, in transiting exoplanets detection, the occulter is assumed to be fully opaque and circularly symmetric, allowing for a fully analytical model \citep{Mandel02}. 
Other model-based approaches to the transit problem may assume, for example, planetary rings, circumstellar discs \citep{Mamajek12}, or dust sheets \citep{Rappaport19, Rein19}. 

However, the general light curve inversion problem for stellar occultation, without any specific model, is yet to be solved.
Recently, a model-free approach to the problem was introduced by the seminal work of \citet[][SK19]{sandford19}, where pixelated gray levels, representing the local opacities of the reconstructed shape, were used as the variables of the problem.

\hyperlink{cite.sandford19}{SK19} pointed out
that the solutions for this ill-posed problem suffer from various degeneracies, which may lead to wrong solutions, as evident when these solutions are compared to their simulated counterparts. These degeneracies include, for example, \textit{Arc degeneracy}, which allows replacing opaque pixels with two opposite stellar-sized arcs, and \textit{Stretch degeneracy}, where the light curve of a transiting shape might be solved with stretched or squeezed versions of that shape, transiting at higher or lower velocities, correspondingly. 

Following \hyperlink{cite.sandford19}{SK19}, the present work suggests a new algorithm, \textit{COD} --- Compact Opacity Distribution, a model-free approach to the light-curve inversion problem, that assumes full opacity for the eclipsing body. \textit{COD} uses a few additional assumptions about the shape simplicity of the occulter, which can apply, for example, to dense clumps of matter in circumstellar disks, cometary-like bodies, or planets with large and tilted rings.
These assumptions allow {\it COD} to construct a model of the eclipsing body with more pixels than the number of data points of the light curve, by significantly reducing the total number of degrees of freedom in the image space, through various constraints and an appropriate objective function.

To significantly narrow the solution space, while keeping it consistent with realistic shapes of large-scale celestial bodies, we apply topological optimization to the occulter’s geometry, 
Our approach consists of two stages. 
In the first stage, \textit{COD} finds the most tightly bound solution,
with all of the opacity-representing pixels lumped together in the most compact arrangement. Then, in the second stage, a few sets of constraints are added to the optimization problem, which ensure monotonicity, continuity and approximate convexity of the occulter's shape, as expected in realistic celestial bodies. The solutions of these two stages depend on two external parameters --- the velocity of the occulter and an auxiliary point, relative to which distances within the occulter's image are minimized.  \textit{COD} estimates these parameters through a systematic search for the values which produce the highest similarity between the results of the two stages.

This work shows that one can formulate the main constraints of the problem, including the topological optimization of the occulter’s geometry, 
by linear relations. This  enables the usage of linear programming (LP) optimization as the mathematical basis of the proposed method. LP is a powerful mathematical tool, which enjoys a fast polynomial convergence while ensuring a {\it global} optimal solution, given there is a feasible one \citep[e.g.,][]{Danzig97, Gonzaga89, Vanderbei08}. 
The global optimality 
effectively narrows the solution space from a highly under-constrained problem into a solvable one, 
enabling to handle a very large number of free variables and constraints. 

Since many practical problems can be expressed as LP problems, it is widely used for planning and optimization in various fields, such as networks, economy, logistics and production, with a large number of available algorithms to solve them \citep[e.g.,][]{Danzig97, Porta00}.
This work uses the Gurobi optimizer as the LP solver, and MATLAB as the programming framework.

Section \ref{section:Mathematical Framework} presents the mathematical framework and formulation of the problem.
In Section \ref{section: Stage I - A Basic Minimization Problem} we develop a preliminary minimization problem, $LP_1$, and discuss its main features and degeneracies, while in Section \ref{section: Stage II - A Further Constrained Minimization Minimization Problem} we present a further constrained problem, $LP_2$,
which prefers simple shapes of opacity distributions.
Section \ref{section:Testing LP_1 and LP_2} performs multiple tests of $LP_1$ and $LP_2$ through simulated transits of fully opaque eclipsing objects with convex contours.
Section~\ref{section:Solving the complete problem}
presents \textit{COD}, the final algorithm, which combines both  $LP_1$ and $LP_2$, followed by tests in Section \ref{section:Testing COD}.
Section~\ref{section:VVV-WIT-08} applies \textit{COD} 
to the mysterious case of VVV-WIT-08 --- a single deep occultation (\( \sim 96 \% \)) of a giant star (\( \gtrsim 30 R_\odot \)),
lasting for over 200 days, resulting in an estimate of the size, velocity and projected opacity distribution of the occulting object.
%
%
Finally, Section~\ref{section:Summary and Discussion} presents a summary of the work and suggests some possible ways forwards. 

\newpage
\section{Mathematical Framework}
\label{section:Mathematical Framework}

Our formulation of the transit light-curve inversion problem as an optimization problem follows \hyperlink{cite.sandford19}{SK19} approach, where the occulter and the host star are represented as pixelated rectangular images, or frames, as plotted in Figure~\ref{fig:framesExample}. 
The frames represent, through gray-scale images, the opacity of the eclipsing body --- {\it{occulter frame}} --- and the stellar flux distribution --- {\it{stellar frame}}. 
The origin of coordinates follows the stellar position on the sky, while the frame of the occulter advances in front of the star, one pixel at each time step throughout the transit, covering different parts of the star.

To keep the pixelization of the two frames consistent with the observed time series, the light curve is interpolated such that each time step corresponds to the motion of the occulter by one stellar pixel.
This framework allows us to represent the gray-scale levels of the occulter frame as the variables of the problem, and to choose a desired function to optimize under various constraints.

\subsection{Setting Up the Two Frames }
\label{section:The Two Frames}
\addcontentsline{toc}{subsection}{The Two Frames}

\begin{figure}
\centering
\includegraphics[scale=1.5]{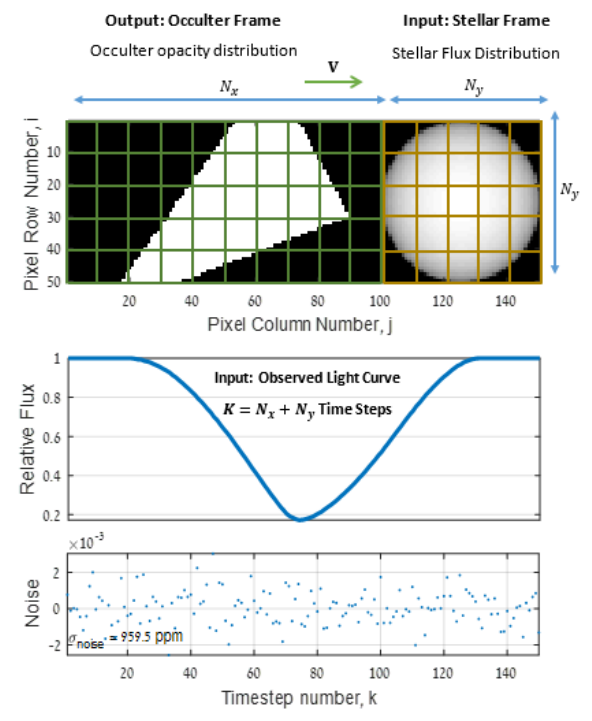}
\caption{{\it Top} –-- Occulter (green grid) and stellar (brown grid) frames. The occulter frame, of size \( N_{x} \times N_{y} = 100 \times 50 \) pixels, advances by \( K = N_x + N_y = 150\) time steps (one time step for each horizontal pixel), until crossing the entire stellar frame. The conversion between time steps and transit times is calculated through the projected relative velocity of the occulter, $V$, and the dimensions of the problem. Here, the light curve is created by the entire frame's occultation, and not just by the opaque object within it. Correspondingly, we aim to reconstruct the entire frame. To be consistent with this mathematical setup, the observed light curve measurements are interpolated to fit the pixels' discretization.
{Middle and bottom} --- the resulting light curve and the simulated noise added to it, correspondingly.}

\label{fig:framesExample}
\end{figure}

In the analysis, we use a frame size of \( N_{y} \times N_{y} \)  pixels (Figure \ref{fig:framesExample}, brown grid), for the stellar frame, and a given model for the stellar flux distribution. This model is integrated over the area of each pixel in the stellar frame, such that \( F_{l,m} \)  represents the total flux emitted by the star within the boundaries of pixel  \(  \left( l,m \right)  \), multiplied by the spectral response of the telescope. Pixels values are normalized such that  \(  \sum _{l,m}^{}F_{l,m}=1 \), with both indices  \( l,m \)  taking the values  \( 1 \ldots N_{y} \).

The occulter's frame, which we aim to find, is of size  \( N_{x} \times N_{y} \)  (Figure \ref{fig:framesExample}, green grid). In this frame, each pixel value,  \( X_{i,j} \), represents the amount of stellar flux blocked by the occulter at location  \(  \left( i,j \right)  \), where  \( i=1 \ldots N_{y} \),  \( j=1 \ldots N_{x} \),  and  \( 0 \leq X_{i,j} \leq 1 \) . 

For a fully transparent pixel,  \( X_{i,j}=0 \), and for a fully opaque one,  \( X_{i,j}=1 \), where values in-between stand for varying degrees of opaqueness. At a certain instant during the transit, the multiplication of the opacity value of each pixel in the occulter's frame with the corresponding flux of a blocked stellar pixel results in the pixel's contribution to the total decrease in the stellar brightness.

The frame of the eclipsing body moves at some projected velocity \( V \) relative to the frame of reference (see Figure \ref{fig:framesExample}), such that each stellar flux 
measurement during the occultation corresponds to a different position of the occulter's frame in front of the star, with a one-pixel translation of this frame along the \( x \)-axis between every two measurements. Therefore, the projected length of one pixel equals \( V \Delta t \), where \( \Delta t \) is the time step of the calculated light curve.
This gives a total of  \( K \equiv N_{x}+N_{y} \) steps from the time when the occulter's frame touches the left edge of the star until it leaves its right edge,
assuming, without loss of generality, that the motion is from left to right.

Since the pixelization of the stellar and occulter frames 
correspond to the time step of the light curve,
the measurements are 
interpolated into \( K \) points in time. We therefore require that the original time series includes 
more than \( K \) data points, such that the reliability of the analysis is not too compromised.
In other words, the number of points within the eclipse determines the maximum resolution of the two frames.


\begin{figure}
\centering		
\includegraphics[scale=0.75]{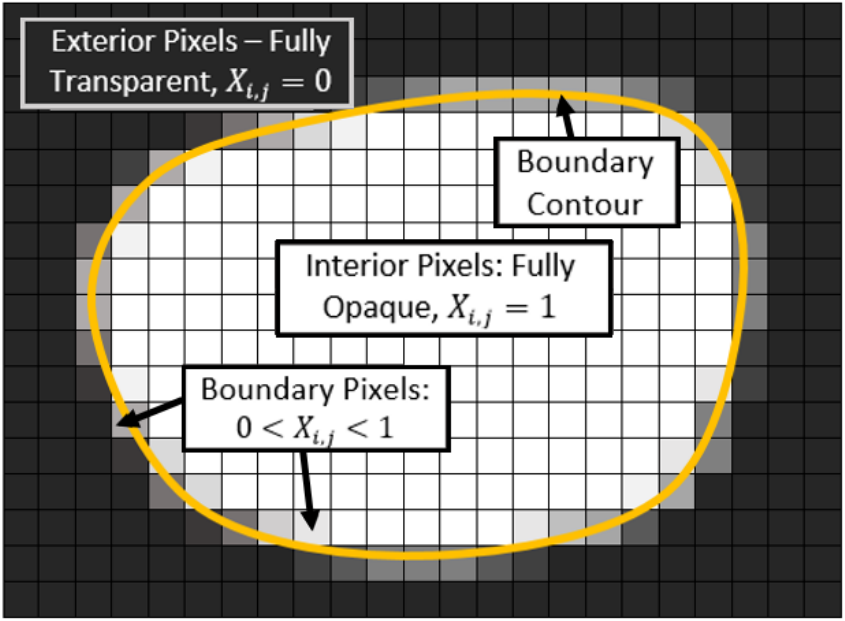}
\caption{A fully opaque occulter: pixels outside the object's boundary are fully transparent (black pixels), whereas pixels well within the boundary are fully opaque (white pixels). Between these regions, at the intersection of the continuous contour of the occulter's shape with the discretized frame pixels, there are pixels with values between $0$ and $1$. The varying gray levels represent the relative area within the pixel that is blocked by the object's edge.}

\label{fig:opticallyThickContour}
\end{figure}

The physical parameters that are known {\it a-priori} in this problem are the total transit time, \( T \), measured from the beginning of ingress to the end of egress, and the time-step between two consecutive measurements, $T/K$. If the stellar radius,  \( R_{\ast} \) is known, then together with the occulter frame's aspect ratio, \( A_{R} \equiv {N_{x}}/{N_{y}} \), one can derive the tangential velocity of the occulter, \( V \), which is assumed --- without loss of generality --- to be aligned with the  \( x \)-axis (by definition of the axes, see Figure \ref{fig:framesExample}):
\begin{equation}
\label{eq:VelocityVsAspectRatio}
V=\left( A_{R}+1 \right) \frac{2R_{\ast}}{T}
=\frac{K}{T}\frac{ 2R_{\ast}}{N_{y}} \ .
\end{equation}

The value of \( N_{y}\), together with the aspect ratio \( A_R\), set the value of \( N_{x}\), and consequently the resolution of our model. Therefore, one would prefer using the largest possible value of \( N_{y}\), while ensuring that \( N_{y}+ N_{x}\) does not significantly exceed the number of independent measurements within the transit.

The following assumptions are taken in our analysis:
\begin{enumerate}
\item The spatial distribution of the stellar flux, \( F_{l,m} \), is known.
\item The \textit{interior} part of the occulter is fully opaque, so it completely blocks the stellar light. When pixelating the occulter into a discrete frame, the pixels outside the boundary of the occulter hold the value of zero opacity (free space) and those inside the boundary hold the value of unity. The pixels on the shape boundary may obtain any value between zero and one, due to the mixing of free space and full opacity (see Figure \ref{fig:opticallyThickContour}).

\item The occulter has a convex shape, i.e., for every two fully opaque pixels in its frame, the pixels along the straight line connecting their centers are fully opaque, too.

\item The occulter does not change its orientation or velocity relative to the stellar frame of reference throughout the entire transit.\par

\end{enumerate}

\subsection{The light curve and the resulting constraints}
\label{section:Formulating an optimization problem}
\addcontentsline{toc}{subsection}{Formulating an optimization problem}

The problem now has  \( N \equiv N_{x} \times N_{y} \)  unknown variables --- all of the pixels opacities in the occulter frame, but only \( K=N_{x}+N_{y} \)  equations: the brightness measurements corresponding to the relative flux at each of the \( K \) steps.
Therefore, the problem is highly under-constrained, and cannot be solved without further assumptions. As a first step, the algorithm aims to significantly narrow the space of possible solutions by the formulation of an optimization problem. The optimization minimizes the value of some topological functions of the opacity distribution, while ensuring that the solution frame reproduces the observed light curve, up to some allowed noise level, through a set of constraints.


The constraints are formulated through the linear relationship between the pixels' opacities \( X_{i,j} \) and the total block of the stellar flux at time step  \( k \). Our goal is to find the frame matrix \( \mathbfss{X} \)  with elements  \( X_{i,j} \), such that it follows the light curve equations –-- i.e. a solution  \( \mathbfss{X} \)  of size  \( N_{x}  \times N_{y} \)  that corresponds to the level of light block at each time step. 

Let \( n \left( i,j \right) \)  be the index of the vectorized form of the pixels matrix, i.e. we rearrange the pixels column by column, into one column vector \( \mathbfit{X} \) with \( N_{y}\times N_{x} \) elements \( X_{n \left( i,j \right) } \). We define  \( A_{n \left( i,j \right),k} \) as the relative stellar flux which pixel \( n \left( i,j \right)\) intersects at time-step  \( k \), such that it is constructed of multiple one-dimensional convolutions, along the \( x \)-axis, of the stellar frame matrix  \( \mathbfss{F} \) with the \( N_{x} \cdot N_{y} \)  elements of the unity matrix, resulting in a total of  \(  \left( N_{x}\cdot N_{y} \right)  \times  \left( N_{x}+N_{y} \right) =N \times K \) elements in the matrix \(\mathbfss{A}\).

The sum
\(  \sum _{n=1}^{N}A_{n \left( i,j \right),k}X_{n \left( i,j \right) } \)  
is the resulting flux decrease at time-step  \( k \), and  \( \mathbfss{A}^{T}\mathbfit{X} \)  gives the entire light curve for the transit of the occulter's frame. \( B_{k} \) is the measured relative flux decrease, or the total light block at time step  \( k \), such that noiseless observations of a pixelated occulter result in  
\( B_{k}= \sum _{n=1}^{N}A_{n \left( i,j \right),k}X_{n \left( i,j \right) }\) for each $k$.

Photometric observations are inherently noisy, so instead of using equalities for the \( N_{x}+N_{y} \) constraints, we use a maximum allowed absolute error margin for each of the \( {K} \) timesteps, \( \varepsilon_{k} \), when fitting the model to the observed light curve: 

\( \vert  \sum _{n=1}^{N}A_{n \left( i,j \right),k}X_{n \left( i,j \right) }-B_{k} \vert  \leq  \varepsilon _{k} ~, \forall k \in  \left\{ 1 \ldots K \right\} \). 

The set of parameters \( \varepsilon_{k} \) can be obtained from the uncertainties of the corresponding measurements.
The value of each \( \varepsilon_{k} \) can be taken, for example, as a few standard deviations of the noise level at each light-curve point. 
Furthermore, one can iteratively decrease all  \(  \varepsilon _{k } \)'s, until the last feasible solution is found.

Note that this method is different from the frequently used $\chi^2$ approach. Instead of minimizing the {\it sum} of (squared) differences between a model and the observations, scaled by their uncertainties, we \textit{constrain} each of the differences to be lower than some maximum value, while minimizing a topological function.

\subsection{The Topological function}

The goal now is to find a topological function to optimize --- a function which its optimal value \textit{prefers} tightly bound and fully opaque convex shapes. 
Let us derive a topological function \(Z\) of the occulter matrix \( \mathbfss{X}\), relative to
some \textit{auxiliary point} \( P \).
We define the vector connecting the pixel  \( n \left( i,j \right)  \)  to \( P \)  as  \( \mathbfit{r}_{n \left( i,j \right) }^{P} \), and $Z$ --- the function to minimize --- as the sum of {\it weighted} distances with respect to that point: 
%
\begin{equation}
 Z \left( \mathbfss{X},P \right) =~ \sum _{n=1}^{N}w_{n \left( i,j \right) } | \mathbfit{r}_{n \left( i,j \right) }^{P} | ,
\end{equation} 
%
where \( | \cdot | \) is the 2-norm. 
The weights \( w_{n \left( i,j \right) } \), which depend on the value at each pixel, are found as follows:
\begin{itemize}
    \item[(i)] Derive the maximum possible contribution of the pixel  \( n \left( i,j \right)  \)  to the light curve,  \( W_{i} \)  (i.e. when assuming that this pixel is fully opaque), which is essentially the discretized integral of the stellar flux along the strip at row  \( i \) :  \( W_{i}= \sum _{m=1 }^{N_{y}}F_{i,m} \).
    \item[(ii)] Find the relative contribution of that pixel to the light curve --- \( w_{n \left( i,j \right) } \) --- through \( W_{i}\) and its individual opacity level: \( w_{n \left( i,j \right) } \equiv W_{i}X_{n \left( i,j \right) } \). 
\end{itemize}  

We now have all the ingredients to construct a minimization problem to find the opacity distribution  \( \mathbfss{X} \) that minimizes \(Z\), while obeying the observational constraints through  \( \varepsilon _{k} \), \(LP_{1}\), 
presented in Section \ref{section: Stage I - A Basic Minimization Problem}. In Section \ref{section: Stage II - A Further Constrained Minimization Minimization Problem}, a further constrained minimization problem is constructed, \(LP_{2}\), which restricts the solution to topologically-preferred shapes.

In both problems, the auxiliary point  \( P \)  and the aspect ratio \( A_{R} \) are given as prior inputs. Since both of these parameters may have a significant impact on the solution, one needs a method of carefully choosing their values. This is done in Section \ref{section:Solving the complete problem}, where \(LP_{1}\) and \(LP_{2}\) serve as sub-problems in the final {\it{COD}} algorithm, which solves the light curve inversion problem without the two prior inputs.
\section{Stage I: A Basic Minimization Problem --- \(LP_{1}\)}
\label{section: Stage I - A Basic Minimization Problem}

At this preliminary step of {\it COD} algorithm, we assume that the 
occulter's projected velocity --- from which we infer the aspect ratio of the occulter's frame --- is known. We also treat the auxiliary point \(P\) as a given input to the algorithm, whereas its geometric meaning is discussed in Section \ref{section:Testing LP_1 and LP_2}. In Section \ref{section:Solving the complete problem} we provide a way to choose the values  of these two parameters through a systematic search.

Both the function to optimize and the constraints of the problem are linear in the variables of the problem --- the elements of \( X \). Therefore, we may formulate the problem as a linear program, designated here as \( LP_{1} \), through the following equations:
\begin{equation}
\label{eq:objectiveFunction}
Minimize~Z \left( \mathbfss{X}|P,\mathbfss{F},\mathbfit{B} \right) =~  \sum _{n=1}^{N}w_{n \left( i,j \right) } \vert \mathbfit{r}_{n \left( i,j \right) }^{P} \vert 
\end{equation}
subject to:

\begin{equation}
\label{eq:constraint1}
  \vert  \sum _{n=1}^{N}A_{n \left( i,j \right),k}X_{n \left( i,j \right) }-B_{k} \vert  \leq   \varepsilon _{k}~,  \forall k \in  \left\{ 1 \ldots K \right\} 
\end{equation}

\begin{equation}
\label{eq:constraintOnVariables}
0 \leq X_{n \left( i,j \right) } \leq  1,  \forall n \in  \left\{ 1 \ldots N \right\},
\end{equation}

The function to optimize, \( Z \left( \mathbfss{X} \right)  \) in Equation (\ref{eq:objectiveFunction}), can be interpreted as the weighted geometric compactness of the opacity distribution \( \mathbfss{X} \) around the point \( P \), given the modelled stellar flux distribution matrix \( \mathbfss{F} \) and the observed light curve vector \( \mathbfit{B}\). 
Equation (\ref{eq:constraint1}) ensures that \( \mathbfss{X} \) produces the observed light curve \( \mathbfit{B} \) up to the maximum allowed difference, and Equation (\ref{eq:constraintOnVariables}) constrains the opacity at each pixel to be physically feasible, between completely transparent and fully opaque. 
\par

The hyper-parameters in  \( LP_{1} \),  are the vertical resolution \( N_{y} \), the aspect ratio \( A_{R} \), the two coordinates of the auxiliary point  \( P \), the maximum allowed error at time step \( k \), \(  \varepsilon _{k} \), and finally --- the stellar flux distribution matrix \( \mathbfss{F} \),
via which \( \mathbfss{A} \) is calculated.

\subsection{Features of the solution} 
\label{subsection:Features of the solution}

\begin{figure}
\centering
\includegraphics[scale=1.1]{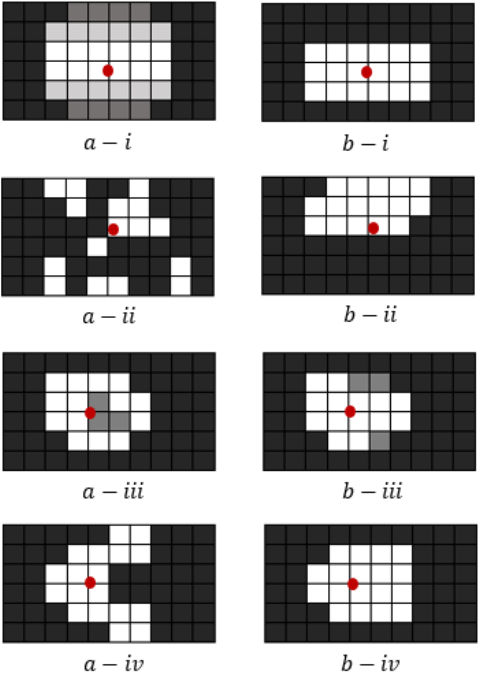}
\caption{\textit{Features of the Solution.} Four properties that the objective function \( Z \) 'prefers' when searching for an optimal solution (column b is preferred over column a): compactness (i), unity (ii), binarity (iii) and convexity (iv). Red dots represent the location of the auxiliary point used in each case.
Given that all of the examples in this figure are feasible solutions that obey the observational constraints (i.e. which recreate the observed light curve up to the allowed errors), the frames in the right column obtain a lower \( Z \) compared with those in the left column. The global optimality feature of LPs ensures in this case that either these solutions are \textit{found} and picked by the algorithm, or a different solution --- which is at least as good --- is chosen.}
\label{fig:3}
\end{figure}

The linearity of both \(Z\) and the constraints ensures that {\it COD} can reach a global optimum, given that there is at least one solution to the problem.
An opacity distribution that \textit{globally} minimizes  \( Z \), while obeying the light curve fit constraints, has four important features, demonstrated in Figure~\ref{fig:3}:

\begin{enumerate}
	\item \textit{Compactness:} A solution containing a compact and tightly bound shape around the auxiliary point \( P \) is preferred over one which is spread over a larger area (Figure ~\ref{fig:3}, first row)
	\item \textit{Unity:} For similar reasoning, a solution containing only one shape is preferred over a solution containing a few separated shapes (Figure ~\ref{fig:3}, second row)
	
	\item \textit{Binarity:} A solution of a fully opaque shape, i.e. with values of unity in the interior pixels and zeros in the exterior ones, is preferred over non-opaque solutions (Figure ~\ref{fig:3}, fourth row)

	\item \textit{Convexity:} Convex solutions are preferred, since \( Z \) benefits from filling cavities --- pixels inside them have shorter distances to the auxiliary point, compared to those farther up, outside the cavities (Figure ~\ref{fig:3}, third row). An analogy would be the sea level which prefers a lower potential energy and globally maintains a convex shape.

\end{enumerate}


\subsection{Solution Degeneracies} 
\label{subsection:Degeneracies in the Solutions}

The {\it Flip} and {\it Stretch} degeneracies, as termed by \hyperlink{cite.sandford19}{SK19}, might still exist in the solutions of the minimization problem.  
The {\it Stretch Degeneracy} relates to solutions having different velocities with corresponding different aspect ratios, which may all solve the \(LP_1\) problem alike. We solve this degeneracy in section \ref{section:Solving the complete problem} through the comparison of the solutions at various aspect ratios and auxiliary points, with a more constrained solution.

The {\it Flip Degeneracy} results from the stellar up-down symmetry around the axis passing through its center and parallel to the occulter's direction of motion --- as projected to the observer. 
The trivial outcome of the degeneracy is that the solution is agnostic to an up-down mirroring of the entire shape around the central axis of the star in the direction of motion.

However, the {\it Flip Degeneracy} allows each column of the solution for the occulter to be flipped up-down independently, while still solving \(LP_1\). What actually limits this set of solutions is our requirement for the solution to remain compact
(see Section \ref{subsection:Features of the solution}). 
We may therefore expect this column-flipping only in significant portions of the occulter at once, and not in separated individual columns. 

\begin{figure*}
\centering
\includegraphics[scale=0.7]{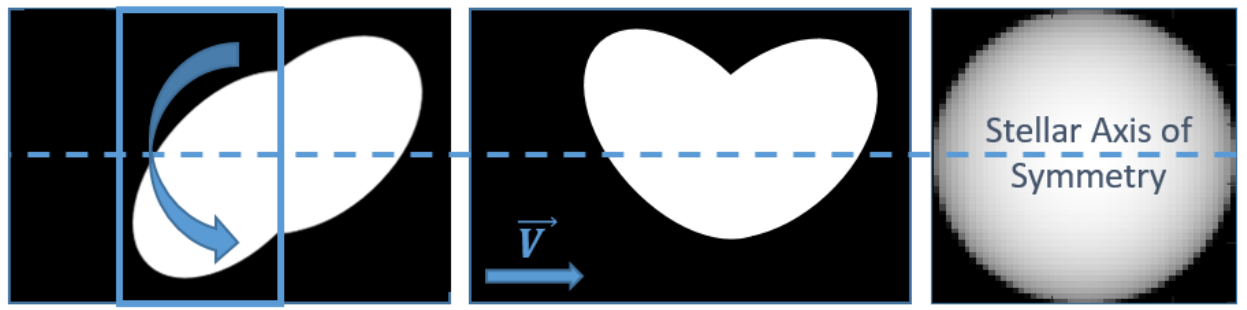}
\caption{{\it Flip Degeneracy} --- an example. {\it{Left:}} The frame of an occulter. The region marked with a box will be flipped up-down around the corresponding stellar axis of symmetry, along the direction of motion (marked with a dashed line). {\it{Center:}} The flipped shape, which obtains exactly the same transit light curve, due to the symmetry of the problem. {\it{Right:}} The flux distribution of the occulted star.}
\label{fig:FlippingExample}
\end{figure*}

An example where only parts of the occulter frame are flipped around the symmetry axis, is presented in Figure \ref{fig:FlippingExample}. The more compact shape, or the one with a lower circumference-to-area ratio, is the one on the left of Figure \ref{fig:FlippingExample}. It is therefore preferred over its flipped variation, as a solution to the problem.


\section{Stage II: A Further Constrained Minimization Problem --- \(LP_{2}\)}
\label{section: Stage II - A Further Constrained Minimization Minimization Problem}

\begin{figure}
    \centering		\includegraphics[scale=0.6]{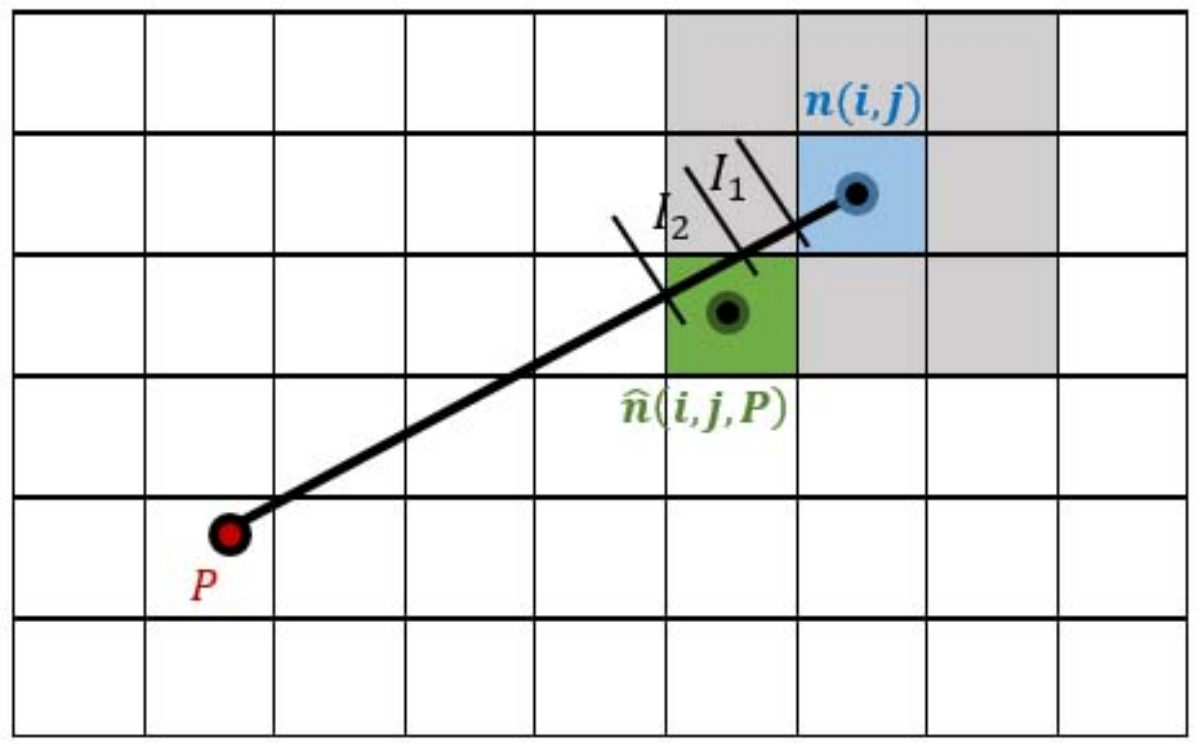}
    \caption{\textit{Approximate radial monotonicity}. To constrain a radial decrease of the pixels values from \( P \) outwards, we first draw a line between \( P~ \) and each pixel in the frame, in this case \( n \left( i,j \right)  \). Then, we find the pixels, among the eight neighbors of \( n \left( i,j \right)  \), which intersect this line. In this case there are two – the pixel to its left, with an intersection length of  \( I_{1} \), and the pixel to its bottom-left, with an intersection length of  \( I_{2} \). Since  \( I_{2}>I_{1} \), we choose the latter one to participate in the constraint, such that  \( X_{n \left( i,j \right) } \leq X_{\hat{n} \left( i,j,P \right) } \). Choosing all intersecting neighbours would have created a chain of undesired constraints, hence only the value of one of them is constrained.
    }
\label{fig:radialMonotonicity}
\end{figure}

\begin{figure}
    \centering		\includegraphics[scale=0.75]{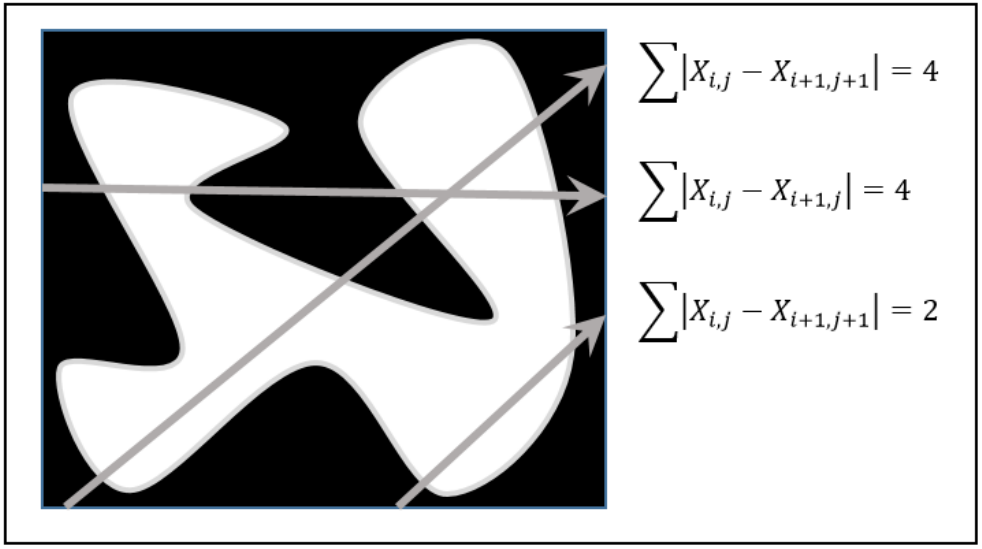}
\caption{\textit{Example: a shape not obeying the approximate convexity constraints}. Cavities inside the shape lead to multiple directional sums --- of the absolute differences in the pixels' opacities --- to be larger than 2. The three sums on the right correspond to the three lines in the figure, counting the total number of changes along them, from 0 to 1 and back. The approximate convexity constraints are not met here, hence this shape cannot be a feasible solution to the problem.} 
\label{fig:ApproximateConvexityExample}
\end{figure}

\begin{figure*}
    \centering		\includegraphics[scale=0.85]{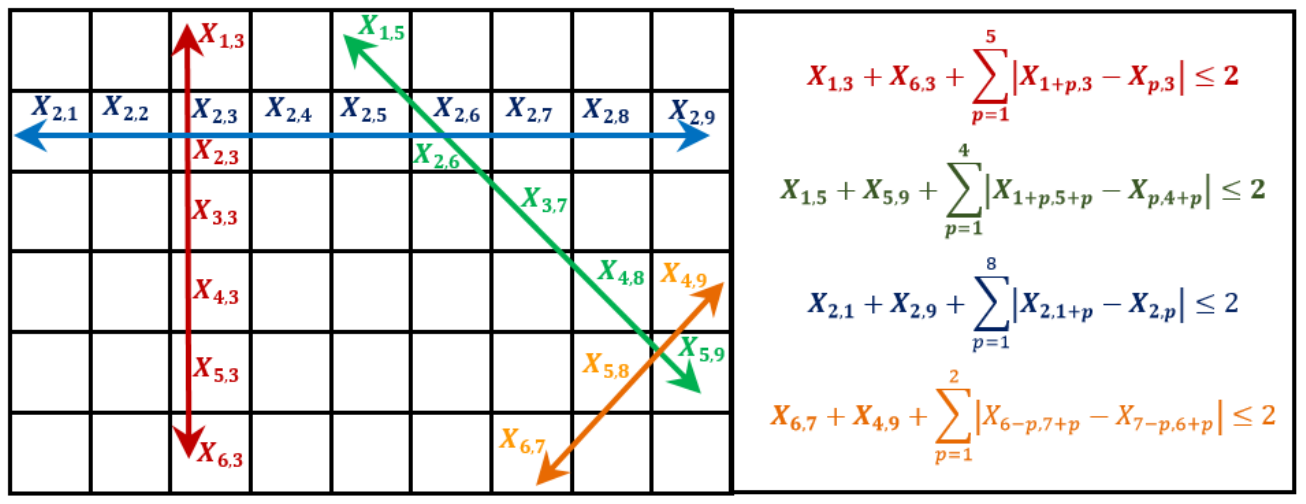}
\caption{\textit{Approximate Convexity Constraints}. The sum of the absolute differences in pixel values along each edge-to-edge line of length \(\geq 3 \), which resides at either the horizontal, vertical or diagonal directions of the grid, is bounded by 2. This allows at most one maximum along each of these lines, and therefore ensures no cavities within the shape's interior along these directions, i.e. partial convexity (up to the discretization of the problem, full convexity would require similar constraints along \textit{all} possible directions).}
\label{fig:ApproximateConvexityContraints}
\end{figure*}

The LP formulation allows adding a large number of linear constraints, without significantly compromising on the computing time needed to reach an optimal solution. We take advantage of this feature to prohibit 
complex occulter shapes, with multiple opacity maxima or non-convex contours. This is done by formulating a further-constrained linear program, \(LP_{2}\),  using two additional sets of constraints:

\begin{enumerate}
\item {\bf Approximate Radial Monotonicity} –-- The opacities of all pixels  decrease from the auxiliary point  \( P \)  outwards. This is achieved by considering the line between  \( P \)  and the center of each pixel  \( n \left( i,j \right)  \) --- the connecting line, and the eight neighbouring pixels of \( n \left( i,j \right)  \) in the solution frame. We
identify the neighboring pixel with the longest intersection with the connecting line, as \( \hat{n} \left( i,j,P \right)  \) --- see Figure~\ref{fig:radialMonotonicity}. Constraining \( \hat{n} \left( i,j,P \right)  \) to be at least as opaque as  \( n \left( i,j \right)  \) ensures a monotonous decrease in the pixels opacity from \( P \) outwards. 

\item {\bf Approximate convexity} –--  
For each pair of pixels, ($i_1,j_1$) and  ($i_2,j_2$), in the frame, the pixel with the lowest opacity along the line connecting them is either ($i_1,j_1$) or ($i_2,j_2$).  
In other words, there are no local opacity minima within the frame, except for its edges. An example of a shape which is {\it{not}} approximately convex appears in 
Figure~\ref{fig:ApproximateConvexityExample}.
In this figure, two of the three lines go through local minima of the opacity, disobeying the approximate convexity requirement. The sum of the absolute differences in opacity along these two lines is $4$, as they include two transitions from $0$ to $1$ and two transitions from $1$ to $0$. 

We therefore require that the sum of the absolute differences in opacity along each line of length  \( \geq 3\), along the grid's horizontal, vertical or diagonal directions, is \(  \leq 2 \) (see Figure \ref{fig:ApproximateConvexityContraints}). 
Note that constraining {\it{full convexity}}, according to this definition, would require implementing this rule along all possible straight lines within the frame, which would be computationally costly, though possible.
\end{enumerate}

Altogether,  \( LP_{2} \) has the same objective function as  \( LP_{1} \) in Equation (\ref{eq:objectiveFunction}), with two sets of constraints, in addition to those of Equations (\ref{eq:constraint1}):
\begin{equation}
\label{eq:constraintRadialMonotonicity}
X_{n \left( i,j \right) } \leq X_{\hat{n} \left( i,j,P \right) },  \forall n \in  \left\{ 1 \ldots N \right\} 
\end{equation}
\begin{equation}
\label{eq:constraintApproximateConvexity}
\sum _{q=1 \ldots L_{ \lambda }-1}^{} \vert X_{q+1}^{ \lambda }-X_{q}^{ \lambda } \vert  \leq  2,  \forall  \lambda  \subset  \Lambda, L_{\lambda} \geq 3 ,
\end{equation} 
where Equation (\ref{eq:constraintRadialMonotonicity}) defines the set of constraints of approximate radial monotonicity from  \( P \) outwards, 
and Equation (\ref{eq:constraintApproximateConvexity}) defines the set of constraints which ensure approximate convexity, with  \( X_{q}^{ \lambda } \)  as the opacity value of the pixel with index  \( q \) along line  \(  \lambda  \subset  \Lambda \), of length  \( L_{ \lambda } \), where \( \Lambda \) is the set of all possible vertical, horizontal or diagonal lines in the occulter's frame.

\section{Testing  \(LP_{1}\) and  \(LP_{2}\) }
\label{section:Testing LP_1 and LP_2}
To test the performance of \(LP_{1}\) and  \(LP_{2}\), we simulated  
a large number of occulters (including physically unlikely shapes with sharp boundaries), generated their transit light curves, and applied the two algorithms to these light curves, similarly to \hyperlink{cite.sandford19}{SK19}. Seven representative simulations are presented here.

For the stellar flux distribution, we assume a quadratic limb darkening law of a sun-like \( G2V \) with solar metallicity star \citep{Claret11}. 
Note that in the case of fast rotators, a gravity-darkening model for the stellar flux should be used \citep{Barnes09}.
The light curves were generated by moving the occulter's frame horizontally pixel-by-pixel and calculating the total relative block of the simulated star’s flux at each time, producing light curves which consist of  \( N_{x}+N_{y} \)  time steps. Random Gaussian noise was added to the light curves at a 
level of  \(1000\) ppm.

All of the simulated occulters are convex and fully-opaque, with a frame size of \( 100 \times 50 \) pixels (\( A_{R}=2\)). Though it would be interesting to analyze the performance of the algorithm for different noise levels, number of measurements during the transit, and frame resolutions, we focus in this work on light curves with relatively low noise levels (\( \lessapprox 1000\) ppm) and large number of measurements ( \( \gtrapprox 100\)), together with occulter frames having a fixed vertical resolution of \(N_{y}=50\) pixels. 

In the LP formulation, we used the noisy light curves as inputs and allowed a maximum absolute deviation of the solutions by \(\varepsilon _{k}=4000 \) ppm, relative all light curve data points \( (1 \leq k \leq K)\).

In our case, where convex shapes are assumed, the original (un-noised) transit light curves are expected to be smooth. Therefore, before solving the LP, we perform a second-degree Savitzky-Golay smoothing \citep{Savitzky64}, with a short span of 5 points.

Here, \( A_{R} \) is taken as the aspect ratio of the simulated occulter, and \( P \) 
as the \textit{weighted geometric 
median}\footnote{See https://en.wikipedia.org/wiki/Geometric\_median} of the occulter's shape. 
This is defined as the point having the smallest sum of distances to all other points in a shape, weighted by the opacity level of pixel \(n(i,j)\), such that:
\begin{equation}
\label{eq:geometricMedian}
P = \argmin_{{P}^{\ast}} \sum _{n(i,j)=1}^{N}X_{n(i,j)} | P^{\ast} - (i,j) |, 
\end{equation}

where the argument \({P}^{\ast} \in \mathbb{R}^2\) may be any point in the plane of the occulter's frame.
The minimum of the expression can be found, for example, through Weiszfeld's algorithm \citep{Chandrasekaran89}.

In the {\it{COD}} algorithm, detailed in Section \ref{section:Solving the complete problem}, the auxiliary point and the aspect ratio are not taken as prior inputs to the solution; the only inputs are the light curve, the maximum allowed deviation of the model, and the stellar model.

Table \ref{tab:testing_LP1_LP2} presents test results for seven simulated occulters. 
The occulting object, the simulated light curve and the resulting optimal frame found by the algorithm, are shown. The last column displays the main metric for the optimal solution, which is the pairwise
(pixel-by-pixel) correlation coefficient \( Corr \).
All correlations are higher than $0.8$, a strong indication for the success of the two algorithms,
which will serve as sub-problems in {\it{COD}}, as described in Section \ref{section:Solving the complete problem}.

The correlations in columns \(4\) and \(6\) of Table \ref{tab:testing_LP1_LP2} indicate that
the further-constrained \(LP_{2}\) generally performed better than \(LP_{1}\). This is the result of the additional constraints in \(LP_{2}\), which aim to restrict the solution to shapes with approximately convex contours.
The advantage of \(LP_{2}\) is used in the last step of {\it{COD}}, which provides the preferred solution to the problem.


\begin{table*}
\centering		
\includegraphics[scale=0.6]{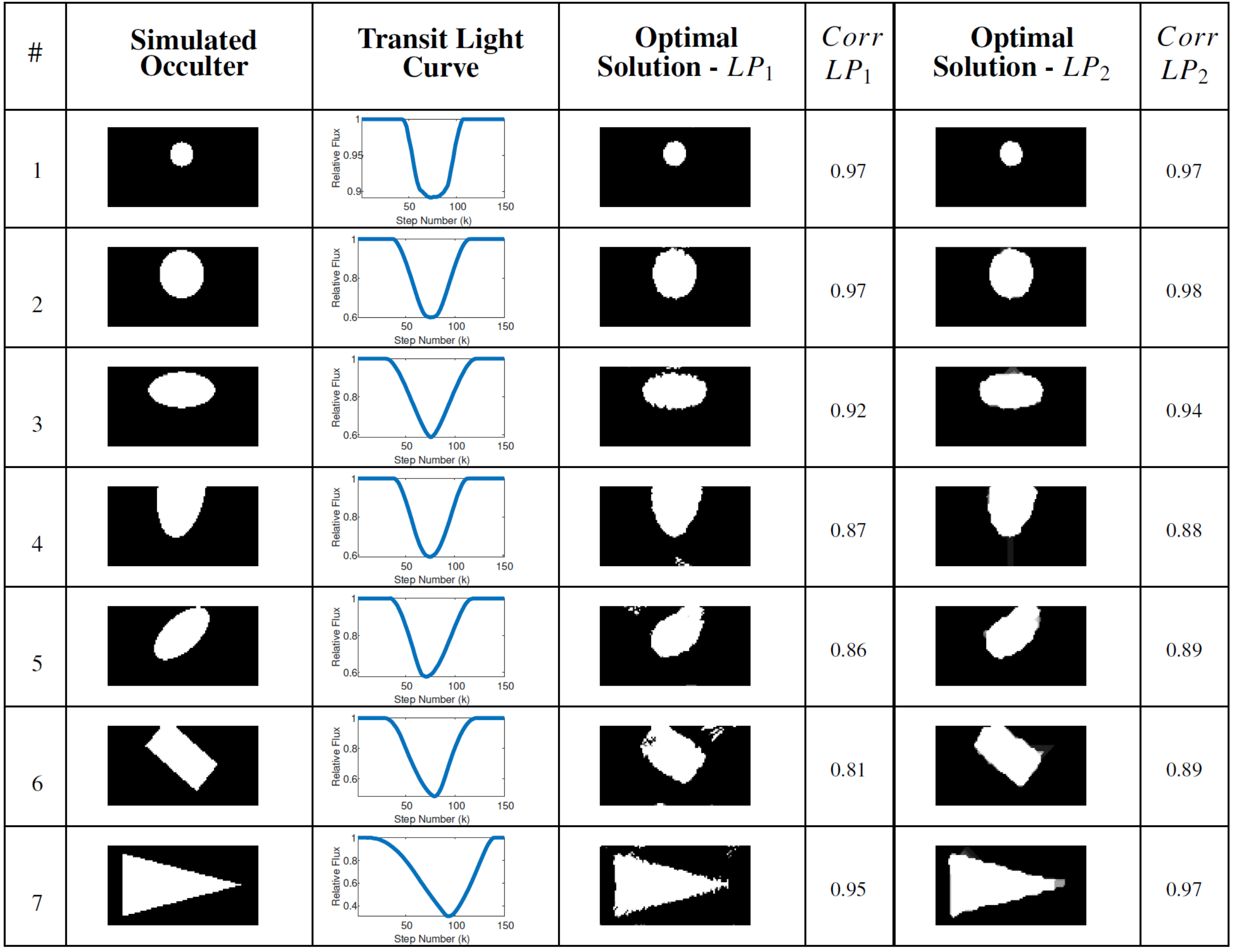}
\caption{
Optimal solutions of the light curve inversion problem, for seven simulated light curves of an eclipse by fully-opaque shapes. 
The frames used were of \(100 \times 50\) pixels. The occulters appear in the first column, and the corresponding simulated light curves, to which a noise level of \(1000 ppm\) was added, appear in column 2. The aspect ratio, \( A_{R}=2 \),  and the auxiliary point \( P \), located in the geometrical median of the occulters, were given as prior inputs to both \(LP_1\) and \(LP_2\). The Solutions of \(LP_1\) and their correlations to the simulated models appear in columns 3 and 4, respectively. The corresponding results of \(LP_2\), to which two sets of constraints were added --- approximate convexity and approximate radial monotonicity --- appear in columns 5 and 6. The solutions of \(LP_2\) are generally better correlated with the simulated occulters than those of \(LP_1\).}

\label{tab:testing_LP1_LP2}
\end{table*}

\section {Solving the Complete Problem via COD}
\label{section:Solving the complete problem}
\addcontentsline{toc}{section}{Solving the complete problem}

In this section, we develop the complete algorithm --- {\it COD} --- which uses both \textit{\( LP_{1} \)} and  \textit{\( LP_{2} \)}
to choose an appropriate location for the auxiliary point \(P\), and find the correct aspect ratio \(A_{R}\).
These parameters, which cannot be extracted directly from the transit light curve, are not be taken at this stage as prior inputs to the solution, as is done in \textit{\( LP_{1} \)} and  \textit{\( LP_{2} \)}.

To understand the role of the two parameters, let us consider the impact of \(P\) and \(A_{R}\) on the large-scale shape of the solution. If \(P\) is located, for example, outside the boundaries of the true shape of the occulter, 
the optimization is expected to result in one of the two: \\
(i) No feasible solution is found, particularly for the more constrained \textit{\( LP_{2} \)}. \\
(ii) Both \textit{\( LP_{1} \)} and \textit{\( LP_{2} \)}
yield a deformed version of the correct shape, to compensate for the poor choice of \(P\). However, \textit{\( LP_{1} \)} has many more degrees of freedom, allowing solutions with non-convex shapes, and is therefore likely to yield in this case a solution that is different from the approximately convex solution of \textit{\( LP_{2} \)}.\\
A similar outcome is expected to result in case \(A_{R}\) is too far from the true aspect ratio of the occulter.

We may try to find the best values of the two parameters through comparing the solutions obtained by \textit{\( LP_{1} \)} and those of the more-restricted minimization problem, \textit{\( LP_{2} \)}. 
Since the two programs use different sets of constraints, the deformation in their solutions for poor choices of \(P\) and \(A_{R}\) is expected to be different. The worse the choices of \(P\) and \(A_{R}\) are, the larger this difference is expected to be. So, to find a solution that is similar in its shape to the correct one, the similarity between the solutions of \textit{\(LP_{1}\)} and \textit{\( LP_{2} \)} may be used as a metric to maximize throughout a systematic search over the values of \(P\) and \(A_{R}\). 

We note that we chose this metric after multiple trials of other metrics. They include, for example, a quantification of the level of convexity of the solution, its surface-to-circumference ratio, the frame's entropy \citep[e.g.,][]{thum84}, its degree of binarity, or its total variation \citep[e.g., first equation in ][]{chambolle04}. 
Maximizing the correlation between \textit{\(LP_{1}\)} and \textit{\( LP_{2} \)} yielded the best results in all our many tests.

The optimal solutions of \( LP_{1} (P,A_R) \)  and  \( LP_{2} (P,A_R) \) are defined here as  \( X_{1}^{\ast} \)  and  \( X_{2}^{\ast} \), correspondingly.
We define the metric \( \hat{C} \left( P,A_{R} \right )  \equiv \textit{corr} \) (\(X_{1}^{\ast} \), \( X_{2}^{\ast} \)), which measures the pixel-by-pixel correlation of the solutions frames of  \( LP_{1} \)  and  \( LP_{2} \). This enables {\it COD} to perform a search on a grid of feasible combinations \(  \left( P,A_{R} \right)  \), and identify the combination \(  \left( P^{\ast},A_{R}^{\ast} \right)  \) which gives the largest  \( \hat{C} \) on that grid, which be used as our final choice for the auxiliary point and the aspect ratio.

{\it COD}'s final solution for the light curve inversion problem is the shape obtained by   \( LP_{2} \)  with $\left( P^{\ast},A_{R}^{\ast} \right)$, as this LP provided better correlations between the simulated occulters and the solutions, as shown in Table \ref{tab:testing_LP1_LP2}.
The impact parameter \( b \) is defined to be the weighted center
of the opacity distribution, and the relative velocity \( V \) is calculated through Equation~(\ref{eq:VelocityVsAspectRatio}), using \( A_{R}^{\ast} \) and the stellar radius.

The search for the optimal \( P^{\ast} \equiv  \left( P_{x}^{\ast},P_{y}^{\ast} \right) \) involves its two coordinates. For a given \( P_{y} \), we found that \( P_{x} \) can be well approximated via a couple of iterations. We first guess the value of \( P_{x} \), and then solve iteratively \( LP_{1} \), where in each iteration, we re-define the  \( x \)-coordinate of the auxiliary point as the \( x \)-coordinate of the weighted geometric median of the of the resulting opacity distribution of the previous iteration, using Equation (\ref{eq:geometricMedian}). Convergence is usually obtained within a couple of iterations.

This iterative process couldn't have worked with regard to \( P_{y} \), due to its strong degeneracy: changing the vertical location of an occulter may be compensated by the deformation of its shape, such that different flux distributions encountered at different stellar cords still reproduce the same light curve.  \( P_{x} \)  does not suffer from such degeneracy, since the horizontal location of the shape relative to the occulter's frame is well defined by the timing of the transit.

While the search for \( P_{y} \) is bounded by the stellar diameter, or here --- the interval \( [0,1] \), \( A_{R} \) is theoretically unbounded.
So, the search for the highest \( \hat{C} \), along the \( A_{R} \) axis, has to have some stopping criterion. Specifically, it is desired to avoid landing on a "stretched" version of the correct solution, one with a higher aspect ratio  (the {\it{Stretch Degeneracy}} in \hyperlink{cite.sandford19}{SK19}), that might produce a higher \( \hat{C} \). 
The search stops when a significant first peak is found --- i.e. it is not a one-point anomaly that may be due to noise, and continues to be the highest value for at least another \( 10\% \) of the \( A_{R} \) axis.

\section{Testing COD}
\label{section:Testing COD}

\subsection{Simulated Occulters}
\label{subsection:Simulated Models vs COD solution}

Similarly to the tests in section \ref{section:Testing LP_1 and LP_2}, the complete algorithm, \textit{COD}, was tested through simulations of multiple transiting fully-opaque convex shapes in front of solar-type stars. We show here the results for the same seven representative simulated occulters as in Section {\ref{section:Testing LP_1 and LP_2}}.
A vertical resolution of \( N_{y} = 50\) pixels is used, following the success of numerous tests of \( LP_1 \) and \( LP_2 \), including the seven tests presented in Section \ref{section:Testing LP_1 and LP_2}.


Table \ref{tab:testing_COD} presents the results of seven simulated occulters similar to those in Table \ref{tab:testing_LP1_LP2}, this time without any prior knowledge on \( A_{R} \)  or  \( P \). Column 3 shows the systematic search through  \( P_{y} \)  and  \( A_{R} \), together with the point \( (P_{y}^{\ast} , A_{R}^{\ast}) \), which yields the maximum value of the correlation \( \hat{C} \) between the solutions of \(LP_1\) and \(LP_2\). Since the values of \( \hat{C} \)  may be very close to 1 around its maximum, we plot \(-log\left(1-\hat{C} \right)\) instead, to better show their contrast, and define it as the \textit{similarity metric} between \(LP_1\) and \(LP_2\). 

The error bounds \( \varepsilon_{k}\) were chosen here to equal approximately \(5\) times the median absolute deviation, calculated from the noise inserted into the light curve, namely: \(  \varepsilon _{k}=4000 \) ppm, for all time steps \(1 \leq k \leq K\). 

Column 4 shows the values of \(P^{\ast}_Y\), together with the solution's impact parameter \( b^{\ast} \) and its distance \( \Delta b \) from the correct one,  \( b \).
Column 5 gives the solution's aspect ratio, \( A^{\ast}_{R} \), together with the resulting error in the projected velocity of the occulter, \(  \Delta V \), which is found through Equation (\ref{eq:VelocityVsAspectRatio}), given the stellar radius. The errors in \( P^{\ast}_{Y} \) and \( A^{\ast}_{R} \) were estimated somewhat arbitrarily, such that they correspond to a decrease of \( 1-\hat{C} \) by \( 50\% \). 
Column 6 presents the shape of the optimal solution, and column 7 gives the bottom-line metric of the algorithm –-- the correlation  \( Corr \) between the solution frame and the simulated frame, pixel-by-pixel. 

The sizes of the resulting frame and the simulated frame might be different, since the aspect ratio (or the occulter's velocity) is unknown. Therefore, to be able to calculate the piece-wise correlation pixel-by-pixel, we interpolated the frame with the higher resolution of the two (or higher \( A_{R} \)) into the lower one.

Altogether, {\it COD} finds feasible solutions to all cases, given only the light curves and the stellar flux models. The algorithm achieves correlations \( \geq 0.87 \) between the solution frames and the simulated frames, with errors in the impact parameters of \(  \Delta b \leq 0.1\ \) and errors in the aspect ratios of  \(  \Delta A_{R}/A_{R} \leq 5\% \). The relative projected velocity is found with relative errors of \(  \Delta V/V \leq 3\% \).


\begin{table*}
\centering		
\includegraphics[scale=0.65]{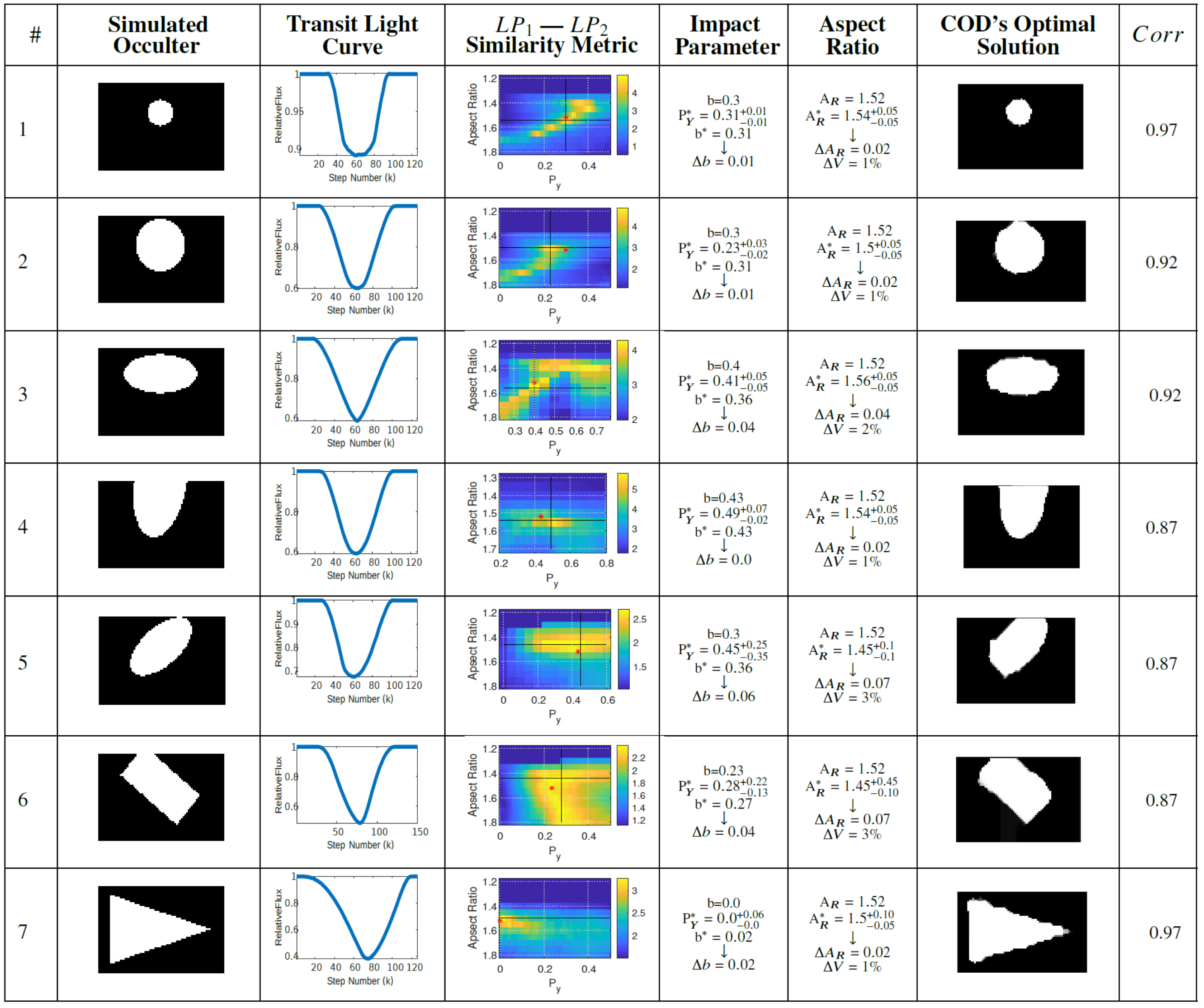}
\caption{Optimal solutions of the light curve inversion problem, using \textit{COD} algorithm, for the same seven simulated occulters and their corresponding noised light curves as in Table \ref{tab:testing_LP1_LP2}. To keep the table compact, a smaller horizontal frame size --- of \( 76 \times 50\) pixels --- was taken.
Column 3 shows the similarity metric, \( -\log  \left\{ 1-\hat{C} \left( P_{y}~,A_{R} \right)  \right\}  \), as a function of \( P_{y} \) and \( A_{R} \). The highest value of this metric is marked with black lines parallel to the two axes, whereas a red star marks the simulated occluter's true aspect ratio and the y-coordinate of its weighted geometric median.
Column 4 shows the calculated value of the impact parameter, \( b^{\ast}\) at the point of highest similarity, \( \left( P_{y}^{\ast}~,A_{R}^{\ast} \right) \), and the difference, \( \Delta b\), between the derived impact parameter and the correct value, \( b\).
Column 5 shows the derived aspect ratio, \(A_{R}^{\ast}\), and its difference \( \Delta A_{R}\) relative to the simulated occulter, \( A_{R} \).
Column 6 presents the optimal solution found, and column 7 shows the bottom-line metric --- the pixel-by-pixel correlation between the solutions and the simulated occulters.
}

\label{tab:testing_COD}
\end{table*}

\subsection{Phobos Eclipse}
\label{subsection:Mars}

\begin{figure*}	
\centering
\includegraphics[scale=0.9]{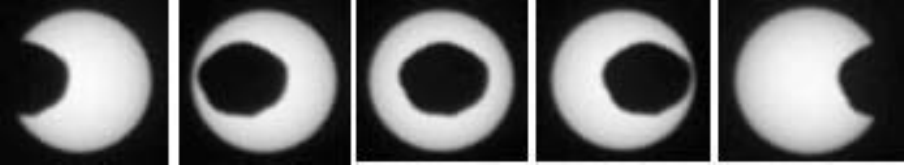}
\caption{Five Snapshots from Phobos' occultation of the Sun, as seen from Mars Curiosity Rover through its telephoto-lens camera, Mastcam. The occultation lasted \(\sim\) 31 seconds, from ingress to egress. The analyzed video was downloaded from \citet[]{nasawebsite}.}
\label{fig:phobosSnapshots}
\end{figure*}

In April 2013, the Mars Curiosity rover has recorded a video of Phobos' occultation of the Sun with a duration of $\sim$ 31 seconds \citep[]{jplwebsite}.
This serves as a good opportunity to test the {\it COD} algorithm, as the event represents a deep occultation (\(\sim 35\% \)) with low noise \( (<1000 \) ppm), a relatively large number of steps during the transit (89 frames) and a convex, yet non-trivial, shape of the occulter (Figure \ref{fig:phobosSnapshots}).

\begin{figure}	
\centering
\includegraphics[scale=0.6]{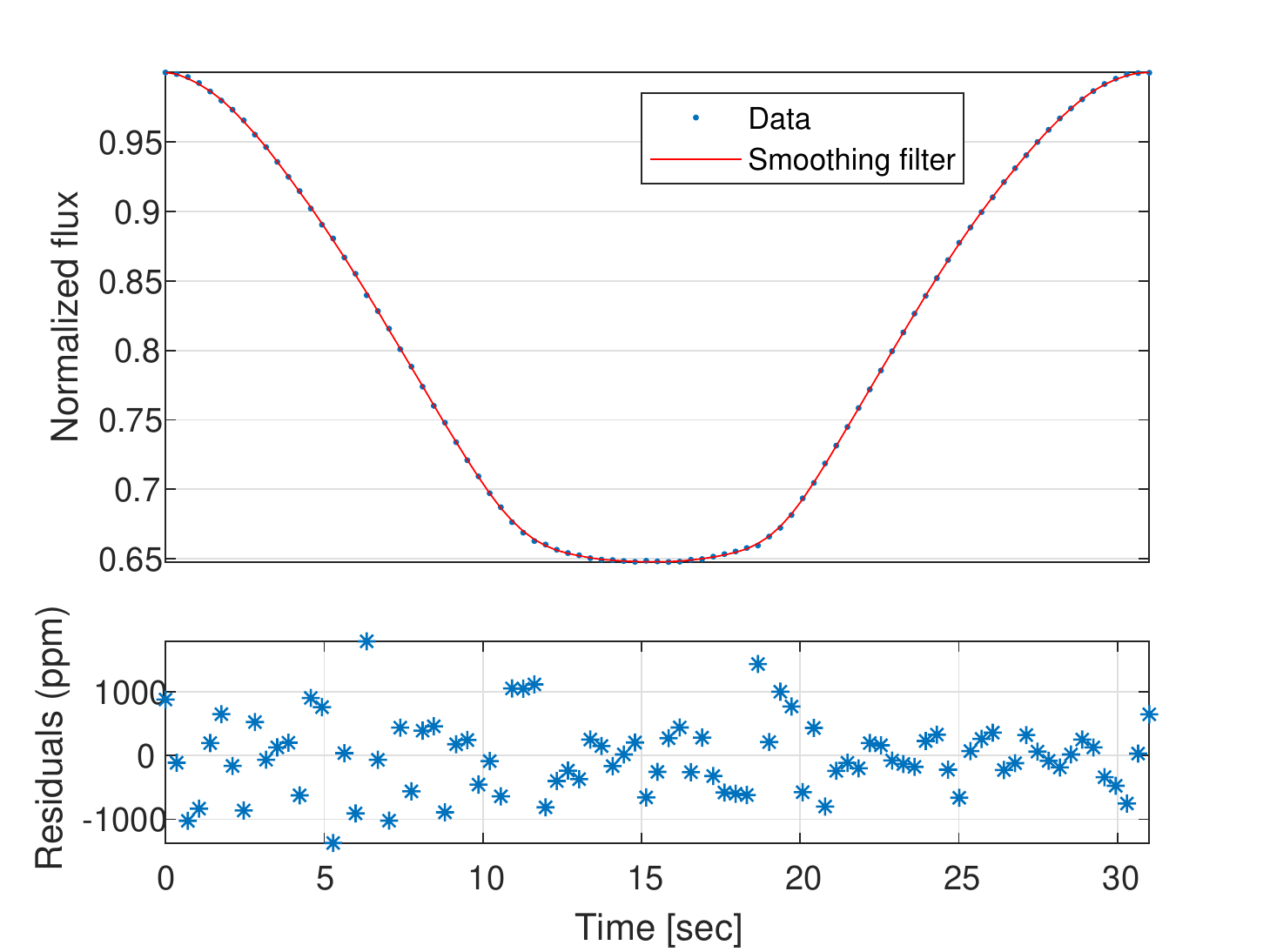}
\caption{Upper panel: the resulting light curve of Phobos eclipse, through the collapse of each frame into one data point, 89 in total. Lower Panel: Deduction of a second-degree Savitzky-Golay smoothing filter from the light curve, with a span of 5 data points. The standard deviation of the residuals, $\sim$ 580 ppm, served as an estimation for the average noise level in the light curve. }
\label{fig:phobosLightCurve}
\end{figure}

To translate the video into a light curve, we summed up all gray levels above a certain threshold in each frame of the video, which was downloaded from \citet[]{nasawebsite}. Unlike the large collecting area, large pixels, high linearity and low overall noise of space telescopes, the Curiosity rover has a modest video camera with a non-linear response at high flux levels \citep[]{Bell17} and a non-Poissonian noise distribution, as was visible in our analysis. 
To separate the bulk of the signal from the added random noise, an identified gray-level threshold of 24 (out of 255) was used. Above this level, all of the pixels' gray levels were summed up into one value. The video came with a batch of 8-12 replications to each frame, with small deviations in the signal levels within these replications. After summing up the gray levels at each frame, we took the median value along each batch of replications, such that we ended up with a total of 89 light-curve values. The resulting light curve is shown in 
Figure~\ref{fig:phobosLightCurve}.

The flux distribution of the Sun, \( F_{k,l} \), was estimated using the Sun's image in the video, at the egress point of the transit. We note that the camera's non-linearity at high fluxes is expected to be a significant source of deformation for the reconstructed shape of Phobos. 
\par

Finally, \textit{COD} algorithm was applied to the light curve using a vertical resolution of \( N_{y}=50 \) pixels, whereas \( N_{x} \) is found through \( A_{R} \).

The resulting reconstructed shape, of size \(55 \times 50\) pixels, is shown in Table \ref{tab:PhobosRestults}, together with the true shape of Phobos' shadow, obtaining a pixel-by-pixel correlation of \(\sim 0.85\). The impact parameter of the solution was found within an error of \(\sim 0.09\). The horizontal angular size of Phobos, relative to the Sun's projected size, \(H\), was found with an error of 
 \(\Delta H/H=9.7\%\).
In contrast to the usual transiting celestial bodies, here the occulter and the star are not located at infinity, so the velocity cannot be inferred from the stellar radius using 
Equation~(\ref{eq:VelocityVsAspectRatio}). However, the error in the relative horizontal angular dimension of Phobos, \( \Delta H \), can be translated to the error in its relative angular velocity \(\Delta  \Omega\), similarly to the logic behind Equation~(\ref{eq:VelocityVsAspectRatio}), yielding \(\Delta\Omega/\Omega=5\%\).
Altogether, {\it COD} succeeded to retrieve the correct parameters of Phobos and its approximate shape.

\begin{table*}
\centering
\includegraphics[scale=0.31]{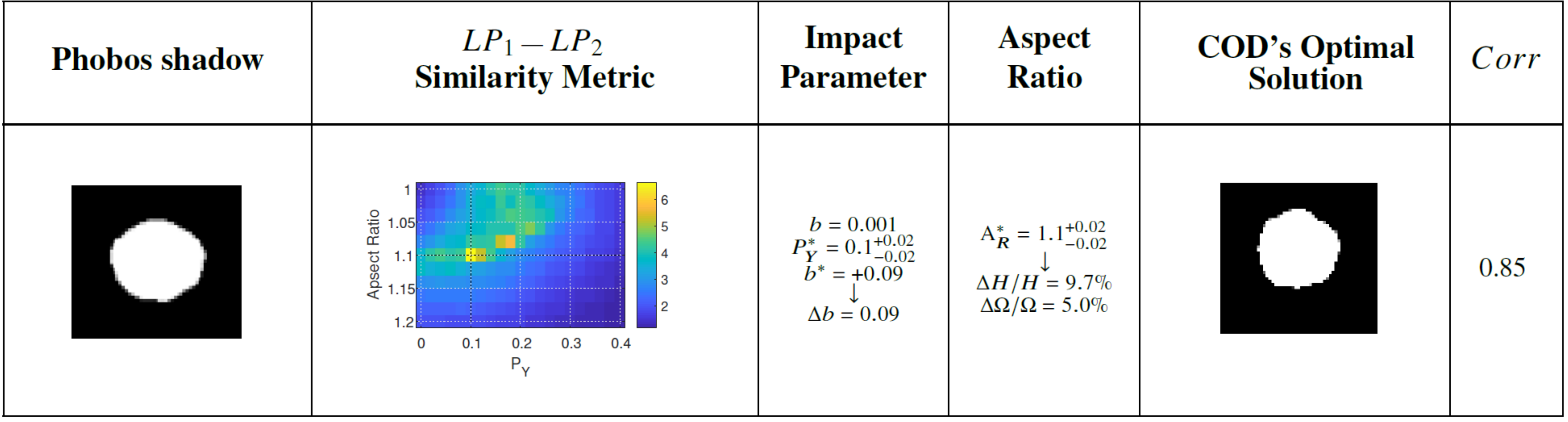}
\caption{The optimal solution of {\it COD} algorithm, using Phobos transit light curve (Figure {\ref{fig:phobosLightCurve}}) as an input, and a vertical resolution of \(N_{y}=50\) pixels. The noise estimate which was used in the solution is \( \sim 5 \) times the standard deviation, leading to \( \varepsilon _{k} \approx 2300 \) ppm \( \forall k \). The error in the impact parameter is \( \Delta b = 0.09 \). 
From the resulting optimal solution, having a size of \( 55 \times 50\) pixels, we can deduce the errors in the horizontal projected angular size of Phobos, relative to the sun's: \( \Delta H/H = 9.7\%\) and consequently the error in the relative angular velocity: \( \Delta \Omega/\Omega = 5\% \). The correlation of the solution's shape with the true one is 0.85.}

\label{tab:PhobosRestults}
 \end{table*}

\section{Application of {\it COD} to VVV-WIT-08}
\label{section:VVV-WIT-08}
\addcontentsline{toc}{section}{The case of VVV-WIT-08}

An eclipse of $\sim96$\% depth of a late-type giant star, VVV-WIT-08, caused by a seemingly opaque object, was recently observed by \citet[][S21]{smith21}, with the Optical Gravitational Lensing Experiment (OGLE) survey \citep[]{udalski15} and the VISTA Variables in the Via Lactea (VVV) surveys \citep[]{minniti10}. The occultation, the only event detected during 17 years of observations of this star, lasted for over 200 days. The light curve strongly suggested a non-spherical shape of the occulter, while, interestingly, was highly achromatic in all observed bands ($V$,$I$ and $K_s$) across the optical and near-infrared wavelengths (\hyperlink{cite.smith21}{S21}).
This suggests an optically thick occulter, which should allow {\it{COD}} to reconstruct its shape, assuming that the shape is simple and convex.
\hyperlink{cite.smith21}{S21} showed that a model of an elliptical projection of the occulter is preferred over a circular one, and calculated the most likely parameters of the ellipse. 

\subsection{Preprocessing}

\begin{figure}
    \includegraphics[scale=0.45]{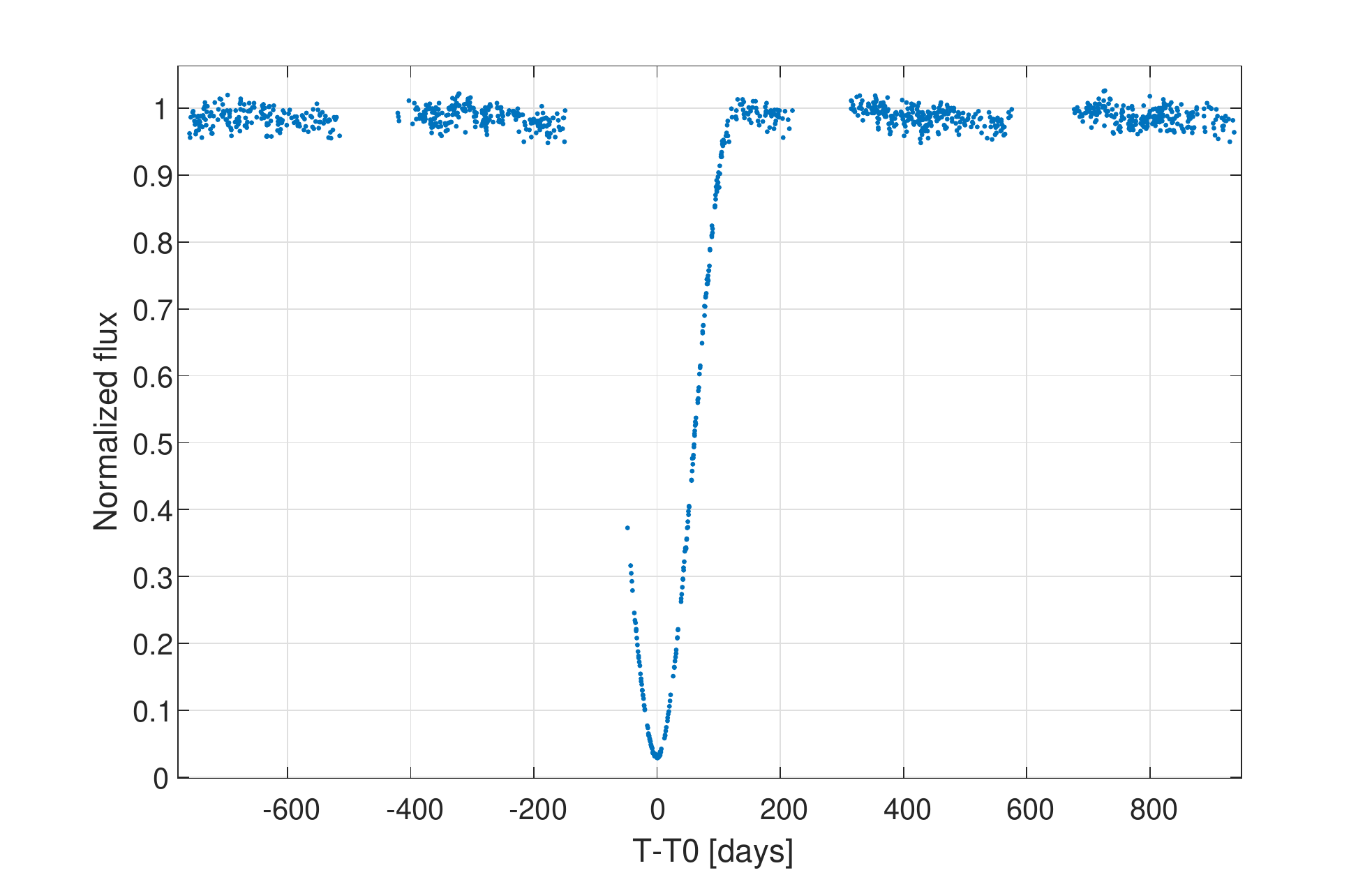}
    \caption{The unprocessed light curve of VVV-WIT-08. The transit contains \(\sim 300\) data points, having a duration of \(\sim 200\) days. A significant part of it is missing, and is therefore extrapolated using the high left-right symmetry, around its deepest point, of the available data points.}
    \label{fig:WITrawLC}
\end{figure}

To apply {\it{COD}}, we used the raw light curve of OGLE $I$ band (Figure \ref{fig:WITrawLC}), which contains the bulk of the data collected during the transit. The data was downloaded from the supplementary material in \hyperlink{cite.smith21}{S21}. The following steps were performed in our prepossessing:
\begin{enumerate}
\item Outliers removal. Out of 313 data points within the occultation, one visual outlier was removed, at \( T-T_0 = 116.7 \) days, where the transit minimum was adopted as $T_0$.
	\item 
	Frequency-domain interpolation. We used the Extended Discreet Fourier Transform algorithm (EDFT) \citep[]{liepins96} to perform a frequency-domain interpolation of the fluctuations in the stellar flux, in and around the transit. EDFT was applied to data within \( \pm 800 \) days around the transit center. The algorithm allows the evaluation of the discrete Fourier transform for incomplete data, in this case --- the purposely removed transit part of the light curve. The latter does not represent the frequency-domain behaviour of the stellar flux, which we are interested in removing during the transit. This interpolation allows the following detrending.
	\item Detrending. The interpolated light curve around the transit was detrended through a least-squares fitting of sines/cosines, with a frequency range of $1/1600$ day$^{-1}$ to $1/100$ day$^{-1}$.

	\item Extrapolation of missing data. In all bands, a significant part of the light curve is missing. The missing data points were extrapolated by mirroring the corresponding ones, around the center of symmetry --- the lowest point of the transit. This is motivated by the very high correlation, \( \sim 99.9\% \), between the available data in both sides of the transit, when mirrored.
	\item Smoothing through a spline interpolation, followed by a Savitzky-Golay filter \citep[]{Savitzky64}, having a span of 12 days in the mid 50 days of the transit, and a more crude span of 24 days in the rest of the transit.
\end{enumerate}

\begin{figure}
   \centering
    \includegraphics[scale=0.62]{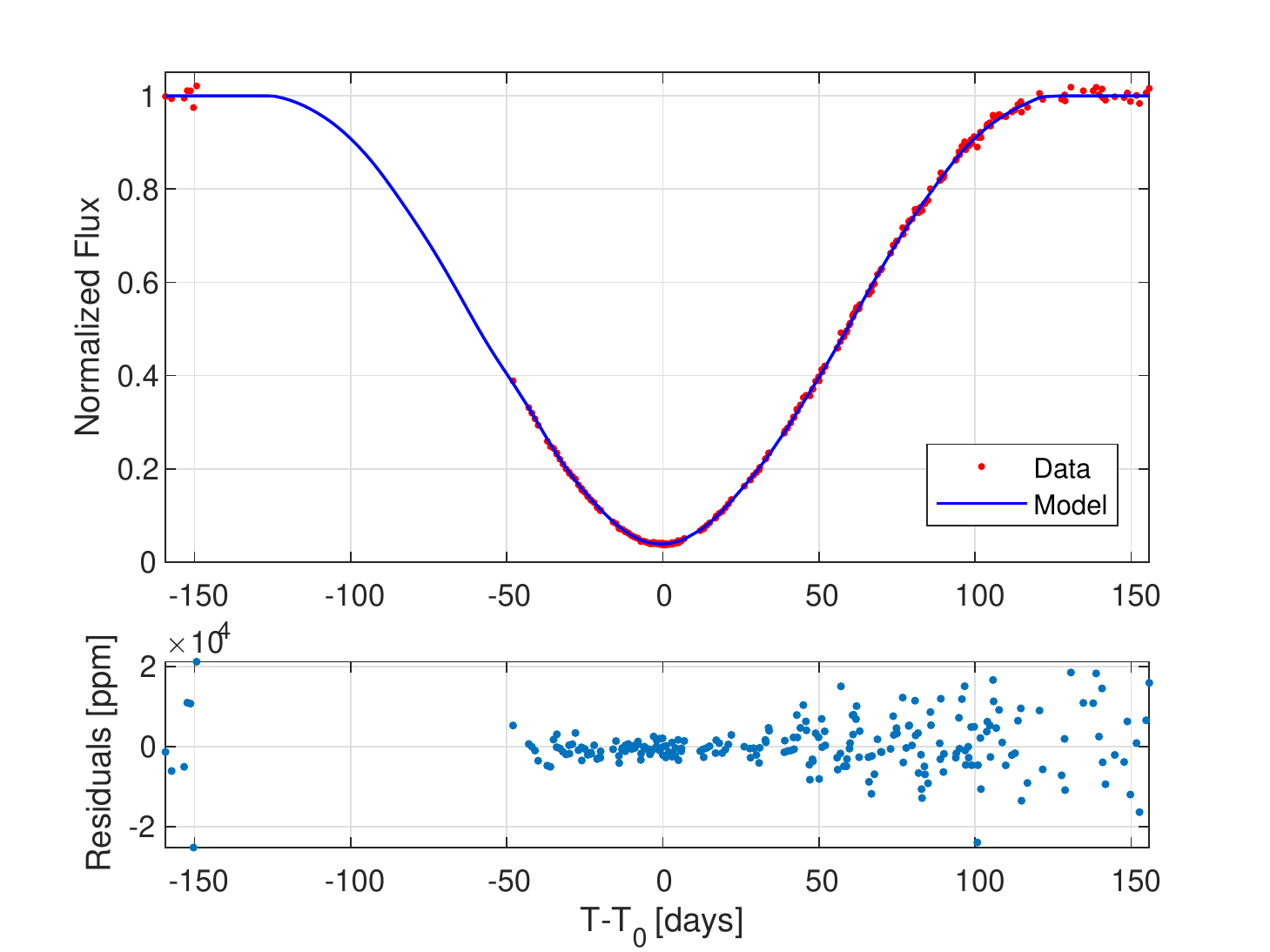}
     \centering
    \caption{Upper panel: red dots - the processed light curve of VVV-WIT-08 during the transit, after outlier removal, EDFT interpolation and detrending. Blue line - the fitted model which was found via smoothing. Lower panel: the residual of the model, having a mean absolute deviation of \( \sim 3500\) ppm.}
    \label{fig:wit08LC}
\end{figure}

The result of the process is presented in Figure \ref{fig:wit08LC}, where the light curve obtained at the end of the pipeline is plotted on top of the detrended data.

\subsection{Applying COD}

As an initial stage of {\it{COD}}, for each aspect ratio in the systematic search along the \(A_R - P_Y\) plane, an interpolation of the processed light curve was created --- as described in Section \ref{section:Solving the complete problem}. 

The stellar radius, which only sets the scale of the problem, was taken according to the "bulge model" of \hyperlink{cite.smith21}{S21}, where a sub-luminous giant star at the distance of the Galactic bulge was assumed, having a radius of \( 30^{+15}_{-10 } R_\odot\). The quadratic limb-darkening coefficients were adopted, too, from \hyperlink{cite.smith21}{S21}. {\it{COD}'s} results were found to be insensitive to variations in the values of these coefficients, within their given uncertainties.

Table \ref{tab:wit08restults} presents the reconstructed shape of VVV-WIT-08, in two extreme scenarios, with and without flux contamination by a nearby faint star of $4$\%, using \( N_{y}=50\) pixels.
For each case we show the results of \( LP_{1} \), alongside the further-constrained \( LP_{2} \), at the point of highest correlation between the two, \( (P_{y}^*,A_{R}^*) \). 

Within the transit, the mean average deviation of the residuals relative to the smoothed model is $\sim 3500$ ppm. The largest absolute deviation which we allowed, \( \varepsilon_{k} \), was five times that value, as in the simulations in Sections \ref{section:Testing LP_1 and LP_2} and \ref{section:Testing COD}. 

Due to the {\it{Flip Degeneracy}} of the problem (see section \ref{subsection:Degeneracies in the Solutions}), we added in Table \ref{tab:wit08restults} an additional step to the process --- an upside-down flip of half the image --- up to the point corresponding to the center of the transit, to which the solution is agnostic due to the stellar up-down symmetry. This leads to an elliptical projection of the occulter, which is more likely than the non-flipped solution, when assuming a gravitationally-bound celestial body, due to its lower surface-to-volume ratio. Note that this case has two axes of symmetry --- one due to the symmetric transit shape, and the other is the stellar symmetry.

The best-fitting ellipses are marked in red in Table~\ref{tab:wit08restults}, and their characteristics are given in Table~\ref{tab:wit08restultsNumeric},
together with those of \hyperlink{cite.smith21}{S21}.
They have eccentricities of \( \sim 0.5 \), are tilted at \( \sim 30^{\circ} \) relative to the direction of motion, travel at $\sim 5$ km/s relative to the host star, and have a semi-minor axis that is similar to the radius of the star. The correlation of the resulting opacity distributions with the best-fitting ellipses are high --- around $95$ \% in both cases (last column in Table \ref{tab:wit08restultsNumeric}).
The two solutions are consistent with those of \hyperlink{cite.smith21}{S21}, who assumed a projected tilted ellipse, unlike our analysis, which did not assume any specific shape of the occulter. 


\begin{table*}
    \includegraphics[scale=0.5]{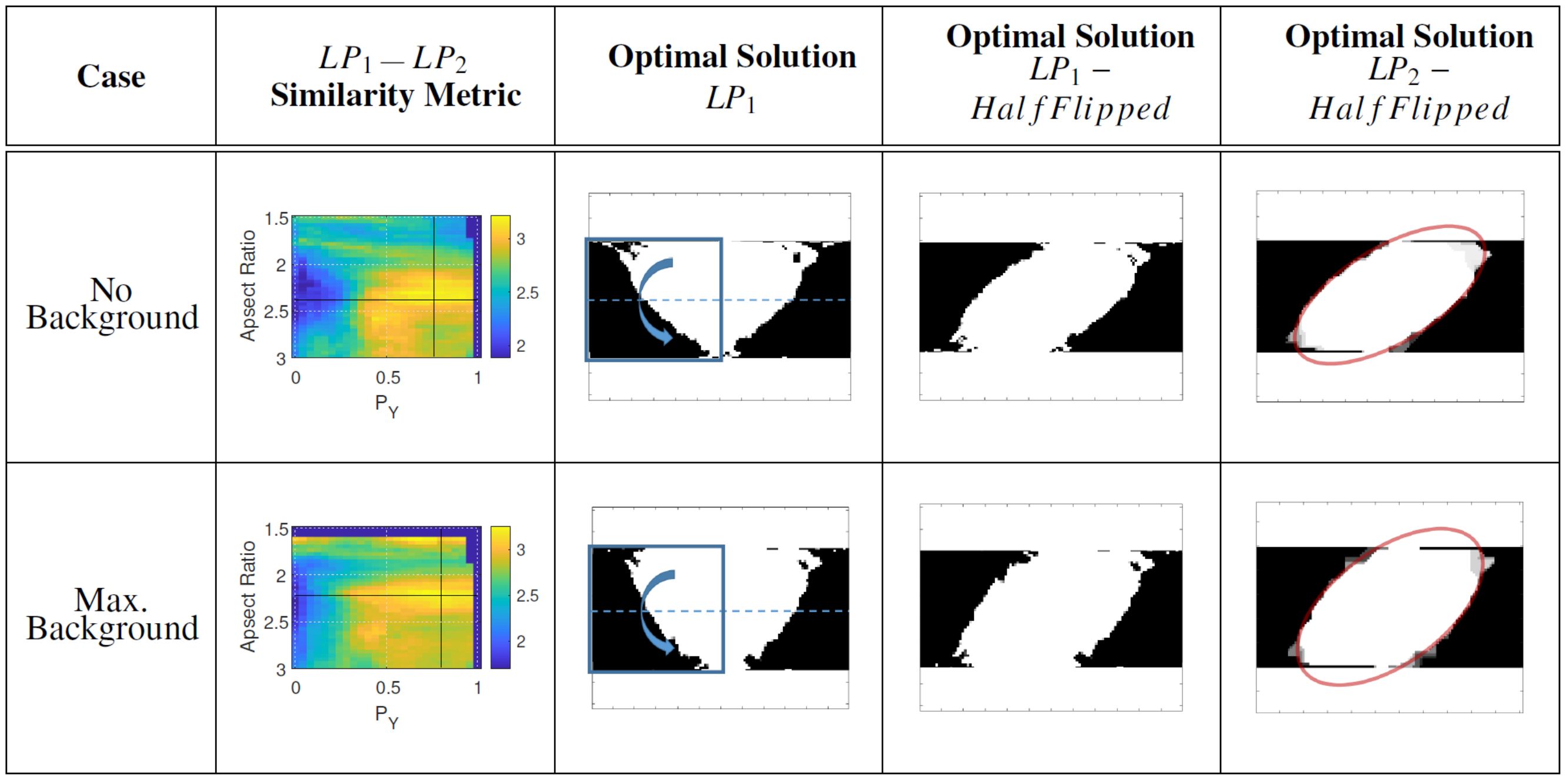}

\caption {{\it COD}'s optimal solution for VVV-WIT-08, using the transit light curve model in Figure {\ref{fig:wit08LC}} as an input, and a vertical resolution of \( N_{y}=50\) pixels. Two cases are presented, with and without  \( \sim 4 \% \) flux contamination.
In addition to the solution of \( LP_1 \), its half-flipped variation is presented too, together with the one of \( LP_2 \). In column 3, the flipped section is marked with a blue box, whereas the axis of up-down symmetry is marked with a dashed blue line.
In column 5, the flipped \( LP_2 \) solutions are presented together with the best-fitting ellipses, which are drawn in red.}

\label{tab:wit08restults}
\end{table*}



\begin{table*}
    \includegraphics[scale=1.25]{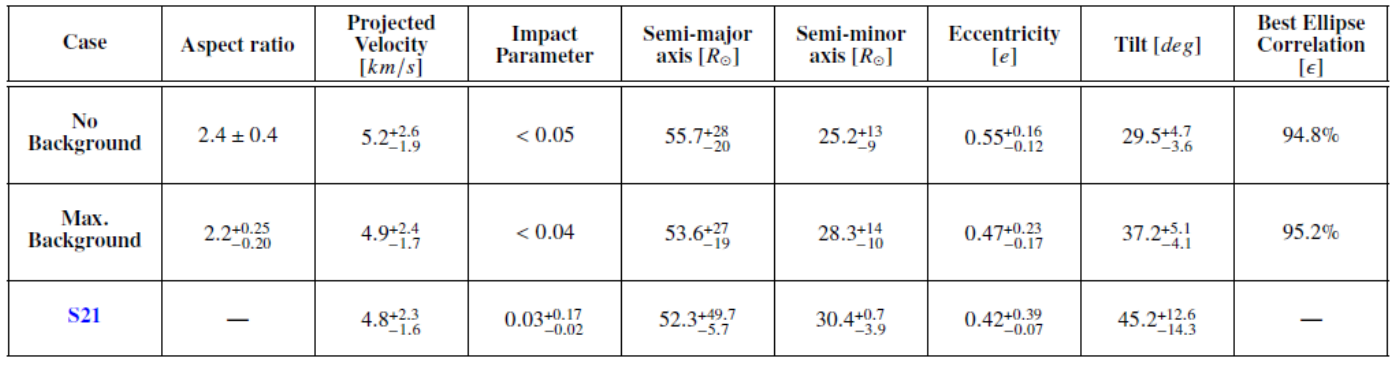}
    
\caption {Characteristics of the best-fitting ellipses of the {\it{COD}} solutions, compared with the 
{\color{blue} S21} solution, which {\it{assumed}} an elliptical shape of the occulter. The main source of errors is the stellar radius, \( 30^{+15}_{-10} R_{\odot}\), as adopted from {\color{blue} S21.} 
}

\label{tab:wit08restultsNumeric}
\end{table*}


\section{Summary and Discussion}
\label{section:Summary and Discussion}
\addcontentsline{toc}{section}{Summary and Conclusions}

We presented a novel algorithm, \textit{COD}, for shape reconstruction of an occulting celestial body, based on a light curve of an occultation event. The algorithm receives as an input a time series of brightness measurements of the eclipsed star during the eclipse and a model for its flux distribution, and returns an estimated opacity distribution of the transiting body, together with its impact parameter and velocity relative to the host star, as projected on the plane of the sky. 
Our goal was to extract as much as possible spatial information of the occulter from the observed light curve, with as few as possible prior assumptions on its shape.

Our main motivation stems from the absence of available technologies for imaging stars and their surrounding celestial bodies with sufficient resolution.
For example, imaging a solar-type star at a distance of $100$ pc with $10\times10$ pixels would require an angular resolution of $ \sim 10\, \mu as $ (micro arc-seconds),
around three orders of magnitudes smaller than the best available angular resolution of existing optical interferometers 
\citep[e.g., \textit{GRAVITY},][]{Abuter17}. Alternatively, earth-sized effective apertures can be obtained through radio telescope arrays, such as the \textit{VLBI}, which achieves an angular resolution of 25$\, \mu as$ 
\citep[e.g.,][]{Akiyama19} through millimeter wave imaging. However, the brightness of main-sequence stars at millimeter-long wavelengths is more than ten orders of magnitude weaker than in the optical band, such that the signal-to-noise ratio is far from the one required for imaging transits.
Therefore, in the foreseen future, the photometric data of extra-solar occultation events will usually yield only a one-dimensional time series of the flux level of the host stars, rather than detailed images.


\textit{COD} is a model-free algorithm for a reconstruction of a fully opaque eclipsing body with an approximately convex contour, being  able to reconstruct the projected shape of the occulter, while using significantly more pixels than the number of data points of the light curve.
We have shown that one can formulate the constraints of the problem, including topological optimization of the geometry of the eclipsing object, by linear relations. We could therefore use a linear programming formulation of an optimization problem, which finds a {\it global} optimum, with a preference to solutions that are approximately convex and optically thick, if they exist. These features correspond to physically feasible solutions of large celestial bodies, tightly held by their own gravitation. Through the comparison of the solutions of two sub-problems, \textit{COD} estimates the impact parameter and the relative projected velocity of the occulter.

The algorithm was successfully applied to many test shapes, including some with sharp-angle boundaries. In addition, we applied {\it COD} to a light curve derived from a video of the 2013 solar eclipse by Phobos, as was recorded from Mars.
In all tests {\it COD} was able to solve the light curve inversion problem, obtaining solutions that yielded high correlations with the original shapes, a strong indication for the success of the algorithm.

Finally, {\it COD} was applied to the intriguing case of VVV-WIT-08 --- an extremely large and optically thick object that has occulted a giant star. Without assuming any specific shape for the occulter, {\it COD} obtained a solution with parameters consistent with those in the discovery paper: a titled opaque ellipse of dimensions similar to the stellar ones, moving at \(\sim 5\) km/s relative to the star. 

The algorithm presented here is a heuristic formulation of some selected topological attributes, without a solid mathematical proof for its general correctness. Nevertheless, the fact that our tests consistently yielded solutions with shapes similar to the simulated ones indicates that the algorithm works well, and provides a first-of-its-kind tool for this type of astronomical data analysis.

A central drawback of \textit{COD}
is its restriction to fully opaque bodies with approximately convex contours. This may include dense gaseous or dusty bodies, but not, e.g., clouds with radially decreasing opacities or rings at high inclinations, with visible inner holes.  
In a follow-up work, we plan to expand the algorithm to enable analyzing occulters having a gradually decreasing opacity, and apply the extended algorithm to the intriguing case of a large asymmetric occultation of \textit{EPIC 204376071} \citep{Rappaport19}, for example. Another effort will focus on expanding the algorithm to include the possibility of hollow shapes through the difference between two fully-opaque concentric ones. This may be the case of the mysterious \textit{Tabby Star} \citep{Boyajian16}, as suggested by \hyperlink{cite.sandford19}{SK19} and \citet{Bourne18}.

Past and present NASA's space telescopes --- \textit{Kepler} \citep[]{Borucki10} and \textit{TESS} \citep[]{Ricker14}, which were designed to find planetary transits, are both great opportunities for the search and analysis of interesting occulters, due to the long uninterrupted light curves they produce, their outstanding sensitivities and lack of atmospheric disruptions. 
In the future, we expect a few projects, 
such as  
PLATO \citep[e.g.,][]{Rauer14},
that are planned to scan the sky photometrically. 
These missions will discover additional stellar eclipses by non-standard bodies, whose shapes may be successfully reconstructed through the \textit{COD} algorithm.

\section*{Acknowledgements}
We deeply thank the anonymous referee for providing detailed comments and suggestions which improved the paper. We also thank Simchon Faigler for fruitful discussions on the EDFT algorithm. This research was supported by Grant No.~2016069 of the United States-Israel Binational Science Foundation (BSF). 
%
\section*{Data availability}
%
No new data were generated or analysed in support of this research.

    
\bibliographystyle{mnras}
\bibliography{LCIbib} 


\bsp	
\label{lastpage}
\end{document}